\documentclass[twoside,12pt]{article}
\usepackage{amssymb}
\usepackage{amsmath}
\usepackage{mathrsfs}
\usepackage{amsfonts}

\begin{document}


\title{ A  boundary matrix for AdS/CFT $SU(1|1)$ spin chain}
\author{  Qingyong Lin$^{1,2}$,  Guangliang Li$^{1,2}$ and Yufei Huang$^{2}$}
\maketitle

\begin{center}
$^1$ {MOE Key Laboratory for Nonequilibrium Synthesis and
Modulation
of Condensed Matter} \\
$^2$ { Department of Applied Physics, Xi'an Jiaotong University,
Xi'an, 710049, China}\\[4mm]
Email:linlqy@stu.xjtu.edu.cn;leegl@mail.xjtu.edu.cn and
yfhuang@mailst.xjtu.edu.cn
\end{center}

\begin{abstract}
  By solving the right reflection equation  proposed in reference\cite{14}  to describe the $Z=0$ giant graviton branes, we obtain
  a boundary matrix with two free parameters for the AdS/CFT $SU(1|1)$ spin chain.
\\[2mm]
  \end{abstract}

{\bf Keywords}: spin chain, reflection equation, boundary matrix


\section{Introduction}

The  discoveries  of integrable structures in planar $N = 4$ SYM
\cite{1y,1,2} and in super- string theory on $AdS_5 \times S^5
$\cite{3,4,5,6}, enable us to determine  this system's spectrum in
the form of a set of non-linear Bethe equations\cite{7,8,8a},
which can be obtained by solving spin chain models
 with different boundaries in the framework of quantum inverse scattering method(QISM)\cite{9,10}.

 In the past years, integrability has been  extended to the open string/spin chain
sector of AdS/CFT\cite{11,12,13}.  Hofman and Maldacena (HM)
 considered open strings attached to maximal giant
gravitons  in $AdS_5 \times S^5$\cite{14}.
 They proposed boundary
S-matrices describing the reflection of world-sheet excitations
(giant magnons) for two cases, namely, the $Y = 0$ and $Z = 0$
giant graviton branes. For the $Y=0$ case, the right boundary
matrix $R^R$ satisfy the reflection equation\cite{14}
\begin{equation}\label{rightbybe1}
    \begin{gathered}
  {S_{12}}\left( {{p_1},{p_2}} \right)R_{1}^R\left( {{p_1}} \right){S_{21}}\left( {{p_2}, - {p_1}} \right)R_{2}^R\left( {{p_2}} \right) \hfill \\
                                = R_{2}^R\left( {{p_2}} \right){S_{12}}\left( {{p_1}, - {p_2}} \right)R_{1}^R\left( {{p_1}} \right){S_{21}}
                                \left( { - {p_2}, - {p_1}}
                                \right),
\end{gathered}
\end{equation}
where $S_{12}(p_1,p_2)$ is the S-matrix determined uniquely by
$SU(2|2)$ symmetry \cite{16}, $R^R$ is a $4\times 4$ matrix and
has a diagonal solution\cite{14,15}.  Murgan and Nepomechie
extended Sklyanin's construction of commuting open-chain transfer
matrices to $Y=0$ case, where they proposed a left reflection
equation and constructed a commuting transfer matrix\cite{17}.
Later the corresponding Bethe equations were obtained by
Galleas\cite{18} and Nepomechie\cite{19} with the help of
different Bethe methods.

For the $Z=0$ case, the giant graviton brane has a boundary degree
of freedom and full $SU(2| 2)$ symmetry\cite{14,15}. The right
boundary S-matrix $R^R$ satisfies the
 right boundary reflection equation (BYBE)\cite{14}
\begin{equation}\label{rightbybe2}
    \begin{gathered}
  {S_{12}}\left( {{p_1},{p_2}} \right)R_{13}^R\left( {{p_1}} \right){S_{21}}\left( {{p_2}, - {p_1}} \right)R_{23}^R\left( {{p_2}} \right) \hfill \\
                                = R_{23}^R\left( {{p_2}} \right){S_{12}}\left( {{p_1}, - {p_2}} \right)R_{13}^R\left( {{p_1}} \right){S_{21}}
                                \left( { - {p_2}, - {p_1}}
                                \right),
\end{gathered}
\end{equation}
where $R^R$ is a $16\times 16$ matrix. Referring the work of
Nepomechie\cite{17,19},  one can prove that if the left BYBE obey
the following equation
\begin{equation}\label{lre}
\begin{gathered}
{S_{21}}{({p_2}, {p_1})^{{t_1}
{t_2}}}R_{31}^L{({p_1})^{{t_1}}}{C_1}( - {p_1}) {S_{21}}{({p_2},
\overline{-p_1})^{{t_2}}}
{C_1}{( - {p_1})^{ - 1}}R_{32}^L{({p_2})^{{t_2}}} \hfill \\
            =  R_{32}^L{({p_2})^{{t_2}}}{C_2}( - {p_2}) {S_{12}}{({p_1},\overline{-p_2})^{{t_1}}}{C_2}{( - {p_2})^{ - 1}}
            R_{31}^L{({p_1})^{{t_1}}}{S_{12}}{( - {p_1}, - {p_2})^{{t_1} {t_2}}},
\end{gathered}
\end{equation}
 the transfer matrix defined as
\begin{equation}\label{transfer}
    \begin{gathered}
  t(p;\{ {q_i}\} )
  =  t{r_{a}}R_{0a}^L(p){T_{a1\cdots L}}(p\ ;\{ {q_i}\} )R_{aL + 1}^R(p) {{\hat T}_{a1 \cdots L}}(p;\{ {q_i}\} ),
\end{gathered}
\end{equation}
can consist of a commuting family, where $C(p)$ is charge
conjugation matrix\cite{20}, $\bar{p}$ denotes the antiparticle
momentum and
\begin{eqnarray}
 {T_{a1 \cdots L}}(p\ ;\{ {q_i}\} )  &=&  {S_a}_L(p, {q_L})  \cdots    {S_{a1}}(p,  {q_1}),\\
{\hat T_{a1 \cdots L}}(p\ ;\{ {q_i}\} )  &=& {S_{1a}}({q_1}, - p)
\cdots  {S_{La}}({q_L}, - p).
\end{eqnarray}
As discussed in reference\cite{18}, the  Hamiltonian for
Eq.(\ref{transfer}) can be defined by
\begin{equation}
H=\frac{d \ln t(p;\{q_i=\pi\})}{dp}|_{p=\pi}.
\end{equation}

 Much work has been done
on the solution to the reflection equation like
Eq.(\ref{rightbybe1})\cite{21,22,23}. However, there is a few work
on the solution to the reflection equation like
Eq.(\ref{rightbybe2})\cite{14,15,24}. For  the AdS/CFT $SU(1|1)$
spin chain\cite{25}, a boundary S-matrix satisfying
Eq.(\ref{rightbybe1}) was found in\cite{17},  while the boundary
S-matrix satisfying Eq.(\ref{rightbybe2}) is not obtained so far
as we know. In this paper we will solve Eq.(\ref{rightbybe2}) to
find the   boundary matrix  for the $SU(1|1)$ spin chain.

 The outline of the paper is organized as follows. In section
2, we  introduce the  $SU(1|1)$-invariant bulk S matrix. In
section 3, we solve Eq.(\ref{rightbybe2}) and present a boundary
matrix with two free parameters. Some discussions are given in
section 4.

\section{Bulk S-matrix of $SU(1|1)$ spin chain}

The graded $SU(1|1)$ bulk S-matrix takes the form\cite{25,17}
\begin{equation}\label{sum1}
S^g(p_1,p_2)=
\begin{pmatrix}
  x_1^+-x_2^- & 0 & 0 & 0 \\
  0 &x_1^--x_2^- & (x_1^+-x_1^-)\frac{\omega_2}{\omega_1} & 0 \\
  0 &(x_2^+-x_2^-)\frac{\omega_1}{\omega_2} & x_1^+-x_2^+ & 0 \\
  0 & 0 & 0 &  x_1^--x_2^+  \\
\end{pmatrix},
\end{equation}
where $x_i^{\pm}$, $\omega_i$ denote $x^\pm(p_i)$, $\omega(p_i)$.
 This S matrix owns unitarity
property and
 crossing-like property\cite{26}.

 Here, for the sake of simplicity, we use the non-graded
$SU(1|1)$ bulk S-matrix
\begin{eqnarray}\label{1}
S(p_1,p_2)=\begin{pmatrix}
  x_1^+-x_2^- & 0 & 0 & 0 \\
  0 &x_1^--x_2^- & (x_1^+-x_1^-)\frac{\omega_2}{\omega_1} & 0 \\
  0 &(x_2^+-x_2^-)\frac{\omega_1}{\omega_2} & x_1^+-x_2^+ & 0 \\
  0 & 0 & 0 & -( x_1^--x_2^+)  \\
\end{pmatrix}
\end{eqnarray}
and let $x^{\pm}({p})$ satisfy the following constraints
\begin{equation}
x^{+}({p})+\frac{1}{x^{+}({p})}-x^{-}({p})-\frac{1}{x^{-}({p})}=\frac{i}{g},\hspace{4mm}
\frac{x^{+}({p})}{x^{-}({p})}=e^{ip}.
\end{equation}
At the same time, we choose $w(p)=1$ due to that
  $w(p)$ can be gauged
away by performing a gauge transformation\cite{26}.

The relation between $S$ and $S^g$ is \cite{17}
\begin{eqnarray}\label{sum2}
S(p_1,p_2)=\mathcal{P}\mathcal{P}^gS^g(p_1,p_2),
\end{eqnarray}
where $\mathcal{P}^g$ and $\mathcal{P}$  are the graded and
non-graded permutation matrix, respectively.
${\mathcal{P}^g}^{ij}_{kl}=(-1)^{p(i)p(j)}\delta_{il}\delta_{jk}$,
$p(1)=0,p(2)=1$ and
${\mathcal{P}}^{ij}_{kl}=\delta_{il}\delta_{jk}$.

\section{Boundary S-matrix of $SU(1|1)$ spin chain}
We now write the right BYBE Eq.(\ref{rightbybe2}) in matrix
element form
\begin{equation}\label{labelbybe}
    \begin{gathered}
  S\left( {{p_1},{p_2}} \right)_{{i_1} {j_1}}^{i j}{R^R}\left( {{p_1}} \right)_{{i_2} {k_1}}^{{i_1} k}S
  \left( {{p_2}, - {p_1}} \right)_{{j_2} i'}^{{j_1} {i_2}}{R^R}\left( {{p_2}} \right)_{j' k'}^{{j_2} {k_1}}  \\
  = {R^R}\left( {{p_2}} \right)_{{j_1} {k_1}}^{j k}S\left( {{p_1}, - {p_2}} \right)_{{i_1} {j_2}}^{i {j_1}}{R^R}
  \left( {{p_1}} \right)_{{i_2} k'}^{{i_1} {k_1}}S\left( { - {p_2}, - {p_1}} \right)_{j' i'}^{{j_2} {i_2}}.  \\
\end{gathered}
\end{equation}
We suppose that the right boundary S-matrix has the same form as
the bulk S-matrix, so we have
\begin{equation}
{R^R}\left( p \right) = \left( {\begin{array}{*{20}{c}}
   {a\left( p \right)} & 0 & 0 & 0  \\
   0 & {b\left( p \right)} & {c\left( p \right)} & 0  \\
   0 & {d\left( p \right)} & {e\left( p \right)} & 0  \\
   0 & 0 & 0 & {f\left( p \right)}  \\
 \end{array} } \right).\label{rm}
\end{equation}
Substituting the boundary matrix Eq.(\ref{rm}) and the bulk
S-matrix Eq.(\ref{1}) into equation (\ref{labelbybe}),  we get 64
equations. we find out that in the 64 equations, there are 48
identical equations. The left 16 equations are the following

\begin{eqnarray}
\nonumber&&  \left( {x_1^ -  + x_2^ + } \right)\left( {x_2^ +  -
x_2^ - } \right){a_1}{e_2} + \left( {x_1^ + + x_2^ + }
\right)\left( {x_1^ +  - x_2^ + } \right){d_1}{c_2} + \left( {x_1^
+  - x_2^ + } \right)\left( {x_2^ +  - x_2^ - }
\right){e_1}{e_2}  \\
&&    - \left( {x_1^ +  + x_2^ - } \right)\left( {x_2^ +  - x_2^ -
} \right){e_1}{a_2} - \left( {x_1^ - - x_2^ - } \right)\left(
{x_2^ +  - x_2^ - } \right){a_1}{a_2} = 0, \label{16-1}
\end{eqnarray}
\begin{eqnarray}
\nonumber&&   \left( {x_1^ -  + x_2^ + } \right)\left( {x_2^ +  -
x_2^ - }\right) {a_1}{d_2} + \left( {x_1^ +  + x_2^ + } \right)\left( {x_1^ +  - x_2^ + } \right){d_1}{b_2}  \\
   &&+ \left( {x_1^ +  - x_2^ + } \right)\left( {x_2^ +  - x_2^ - }\right) {e_1}{d_2} - \left( {x_1^ +  + x_2^ - } \right)\left(
{x_1^ +  - x_2^ - } \right){d_1}{a_2} = 0,  \label{16-2}
\end{eqnarray}
\begin{eqnarray}
\nonumber&&
  \left( {x_1^ -  - x_2^ - } \right)\left( {x_1^ +  - x_1^ - } \right){a_1}{a_2} + \left( {x_1^ +  + x_2^ - } \right)\left(
{x_1^ +  - x_1^ - } \right){e_1}{a_2} - \left( {x_1^ +  + x_2^ + } \right)\left( {x_1^ +  - x_2^ + } \right){c_1}{d_2}  \\
&&   - \left( {x_1^ -  + x_2^ + } \right)\left( {x_1^ +  - x_1^ -
} \right){a_1}{e_2} - \left( {x_1^ +  - x_2^ + } \right)\left(
{x_1^ +  - x_1^ - } \right){e_1}{e_2} = 0, \label{16-3}
\end{eqnarray}
\begin{eqnarray}
&&\left( {x_1^ +  + x_2^ + } \right)\left( {x_1^ +  - x_1^ - }
\right){d_1}{c_2} - \left( {x_1^ +  + x_2^ + } \right)\left( {x_2^
+  - x_2^ - } \right){c_1}{d_2} = 0,\label{16-4}
\end{eqnarray}
\begin{eqnarray}
\nonumber&&
  \left( {x_1^ -  + x_2^ + } \right)\left( {x_1^ -  - x_2^ - } \right){a_1}{d_2} + \left( {x_1^ +  + x_2^ + } \right)\left( {x_1^ +  - x_1^ - } \right)
  {d_1}{b_2} + \left( {x_1^ +  - x_1^ - } \right)\left( {x_2^ +  - x_2^ - } \right){e_1}{d_2}  \\
 &&  - \left( {x_1^ +  + x_2^ + } \right)\left( {x_1^ +  - x_2^ - } \right){b_1}{d_2} - \left( {x_1^ +  - x_2^ - } \right)\left( {x_1^ +  - x_1^ - } \right){d_1}{e_2} = 0, \label{16-5}
\end{eqnarray}
\begin{eqnarray}
\nonumber&&
  \left( {x_1^ +  + x_2^ - } \right)\left( {x_1^ +  - x_2^ - } \right){c_1}{a_2} - \left( {x_1^ +  + x_2^ + } \right)
  \left( {x_1^ +  - x_2^ + } \right){c_1}{b_2}  \\
&&   - \left( {x_1^ -  + x_2^ + } \right)\left( {x_1^ +  - x_1^ -
} \right){a_1}{c_2} - \left( {x_1^ +  - x_2^ + } \right)\left(
{x_1^ +  - x_1^ - } \right){e_1}{c_2} = 0, \label{16-6}
\end{eqnarray}
\begin{eqnarray}
\nonumber&&
  \left( {x_1^ +  + x_2^ + } \right)\left( {x_1^ +  - x_2^ - } \right){b_1}{c_2} + \left( {x_1^ +  - x_2^ - } \right)\left( {x_2^ +  - x_2^ - } \right){c_1}{e_2}
  - \left( {x_1^ +  + x_2^ + } \right)\left(
{x_2^ +  - x_2^ - } \right){c_1}{b_2}  \\
   &&- \left( {x_1^ -  + x_2^ + } \right)\left( {x_1^ -  - x_2^ - } \right){a_1}{c_2} - \left( {x_1^ +  - x_1^ - } \right)\left( {x_2^ +  - x_2^ - } \right){e_1}{c_2} = 0,  \label{16-7}
\end{eqnarray}
\begin{eqnarray}
&&\left( {x_1^ +  - x_2^ - } \right)\left( {x_2^ +  - x_2^ - }
\right){c_1}{d_2} - \left( {x_1^ +  - x_1^ - } \right)\left( {x_1^
+  - x_2^ - } \right){d_1}{c_2} = 0,\label{16-8}
\end{eqnarray}
\begin{eqnarray}
\nonumber&&
  \left( {x_1^ +  + x_2^ - } \right)\left( {x_1^ +  - x_1^ - } \right){f_1}{b_2} + \left( {x_1^ -  - x_2^ - } \right)\left(
{x_1^ +  - x_1^ - } \right){b_1}{b_2} - \left( {x_1^ -  + x_2^ - } \right)\left( {x_1^ -  - x_2^ - } \right){c_1}{d_2}  \\
 &&  - \left( {x_1^ -  + x_2^ + } \right)\left( {x_1^ +  - x_1^ - } \right){b_1}{f_2} - \left( {x_1^ +  - x_1^ - } \right)\left(
{x_1^ +  - x_2^ + } \right){f_1}{f_2} = 0, \label{16-9}
\end{eqnarray}
\begin{eqnarray}
\nonumber&&
  \left( {x_1^ +  + x_2^ - } \right)\left( {x_1^ +  - x_1^ - } \right){f_1}{c_2} + \left( {x_1^ -  - x_2^ - } \right)\left(
{x_1^ +  - x_1^ - } \right){b_1}{c_2}  \\
 &&  - \left( {x_1^ -  + x_2^ - } \right)\left( {x_1^ -  - x_2^ - } \right)
 {c_1}{e_2} + \left( {x_1^ -  + x_2^ + } \right)\left( {x_1^ -  - x_2^ + } \right){c_1}{f_2} = 0, \label{16-10}
 \end{eqnarray}
\begin{eqnarray}
\nonumber&&
  \left( {x_1^ -  + x_2^ + } \right)\left( {x_2^ +  - x_2^ - } \right){b_1}{f_2} + \left( {x_1^ +  - x_2^ + } \right)\left(
{x_2^ +  - x_2^ - } \right){f_1}{f_2} + \left( {x_1^ -  + x_2^ - } \right)\left( {x_1^ -  - x_2^ - } \right){d_1}{c_2}  \\
&&   - \left( {x_1^ -  - x_2^ - } \right)\left( {x_2^ +  - x_2^ -
} \right){b_1}{b_2} - \left( {x_1^ +  + x_2^ - } \right)\left(
{x_2^ +  - x_2^ - } \right){f_1}{b_2} = 0, \label{16-11}
\end{eqnarray}
\begin{eqnarray}
&&\left( {x_1^ -  + x_2^ - } \right)\left( {x_2^ +  - x_2^ - }
\right){c_1}{d_2} - \left( {x_1^ -  + x_2^ - } \right)\left( {x_1^
+  - x_1^ - } \right){d_1}{c_2} = 0,\label{16-12}
\end{eqnarray}
\begin{eqnarray}
\nonumber&&
  \left( {x_1^ +  - x_1^ - } \right)\left( {x_2^ +  - x_2^ - } \right)
{b_1}{c_2} + \left( {x_1^ +  + x_2^ - } \right)\left( {x_1^ +  -
x_2^ + } \right){f_1}{c_2} - \left( {x_1^ -  + x_2^ - }
\right)\left( {x_2^ +  - x_2^ - } \right)
{c_1}{e_2}  \\
&&   + \left( {x_1^ -  - x_2^ + } \right)\left( {x_2^ +  - x_2^ -
} \right){c_1}{b_2} - \left( {x_1^ -  + x_2^ - } \right)\left(
{x_1^ -  - x_2^ + } \right){e_1}{c_2} = 0, \label{16-13}
\end{eqnarray}
\begin{eqnarray}
\nonumber&&
  \left( {x_1^ -  + x_2^ + } \right)\left( {x_1^ -  - x_2^ + } \right){d_1}{f_2} + \left( {x_1^ +  + x_2^ - } \right)\left( {x_2^ +  - x_2^ - } \right)
{f_1}{d_2}  \\
 &&  + \left( {x_1^ -  - x_2^ - } \right)\left( {x_2^ +  - x_2^ - } \right)
{b_1}{d_2} - \left( {x_1^ -  + x_2^ - } \right)\left( {x_1^ -  -
x_2^ - } \right){d_1}{e_2} = 0,  \label{16-14}
\end{eqnarray}
\begin{eqnarray}
\nonumber&&
  \left( {x_1^ -  - x_2^ + } \right)\left( {x_1^ +  - x_1^ - } \right)
{d_1}{b_2} - \left( {x_1^ -  + x_2^ - } \right)\left( {x_1^ -  -
x_2^ + }
\right){e_1}{d_2} + \left( {x_1^ +  + x_2^ - } \right)\left( {x_1^ +  - x_2^ + } \right){f_1}{d_2}  \\
&&   + \left( {x_1^ +  - x_1^ - } \right)\left( {x_2^ +  - x_2^ -
} \right){b_1}{d_2} - \left( {x_1^ -  + x_2^ - } \right)\left(
{x_1^ + - x_1^ - } \right){d_1}{e_2} = 0, \label{16-15}
\end{eqnarray}
\begin{eqnarray}
&&\left( {x_1^ -  - x_2^ + } \right)\left( {x_1^ +  - x_1^ - }
\right){d_1}{c_2} - \left( {x_1^ -  - x_2^ + } \right)\left( {x_2^
+  - x_2^ - } \right) {c_1}{d_2} = 0,\label{16-16}
\end{eqnarray}
where $a_i=a(p_i)$, so do $b_i, c_i, d_i,e_i,f_i$.  The property
$x^{\pm}(-p)=-x^{\mp}(p)$ are used\cite{15}.

 In order to solve $R^R$, we differentiate  the above 16 equations with
$p_2$, and let $p_2=p_0$, then we will get 16 equations about $p$,
from which we can solve $R^R$. During this process, we introduce
the following 16 parameters, $x^+(p_0)$, $x^-(p_0)$, $x^+(p_0)'$,
$x^-(p_0)'$, $a(p_0)$, $a(p_0)'$, $b(p_0)$, $b(p_0)'$, $c(p_0)$,
$c(p_0)'$, $d(p_0)$, $d(p_0)'$, $e(p_0)$, $e(p_0)'$, $f(p_0)$,
$f(p_0)'$. We believe that these 16 parameters will be
self-consistent, which offers the chance to simplify the
equations.

If we  let the initial condition be $R^R(p_0)=I$, where $I$ is
$4\times 4$ identity matrix,
 we will find at once that
 \begin{equation}\label{}
 x^+(p_0)=-x^-(p_0),\quad [x^+(p)]'_{p=p_0}=[x^-(p)]'_{p=p_0}.
 \end{equation}
 Considering that\cite{15}
\begin{equation}\label{}
x^{\pm}(-p)=-x^{\mp}(p),\quad {x^ \pm }\left( p \right) =
\frac{{{e^{ \pm \frac{{ip}} {2}}}}} {{4g\sin\left( {\frac{p} {2}}
\right)}}\left( {1 + \sqrt {1 + 16{g^2}\sin^2\left( {\frac{p} {2}}
\right)} } \right),
    \end{equation}
we find $p_0=(2k+1)\pi$. Without losing generality, we choose
$p_0=\pi$. Then we have
\begin{eqnarray}\label{ic}
\nonumber&&a(\pi)=b(\pi)=e(\pi)=f(\pi)=1,\quad c(\pi)=d(\pi)=0\\
&&x^+(\pi)=-x^-(\pi)=x_B,\quad
x^+(p)'\mid_{p=\pi}=x^-(p)'\mid_{p=\pi}=\frac{i}{2}x_B,
\end{eqnarray}
where ${x_B} = \frac{i} {{4g}}\left( {1 + \sqrt {1 + 16{g^2}} }
\right)$. With the help of Eq.(\ref{ic}), we find that equation
(\ref{16-4}), (\ref{16-8}), (\ref{16-10}) and (\ref{16-16}) are
identical, so do equation (\ref{16-1}) and (\ref{16-3}), equation
(\ref{16-2}) and (\ref{16-6}), equation (\ref{16-5}) and
(\ref{16-7}), equation (\ref{16-9}) and (\ref{16-11}), equation
(\ref{16-12}) and (\ref{16-14}) and equation (\ref{16-13}) and
(\ref{16-15}). Simplifying these equations, we finally obtain one
solution to the right reflecting equation Eq.(\ref{rightbybe2})
\begin{eqnarray}
a& =&  - \frac{{\xi\eta\left( {{x^ + } - {x_B}} \right)\left[
{{{\left( {{x^ + }} \right)}^2} - {{\left( {{x_B}} \right)}^2}}
\right] + 2{{\left( {{x^ + }} \right)}^2}\left( {2{x^ + } - {x_B}}
\right)}} {{\xi\eta\left( {{x^ - } + {x_B}} \right)\left[
{{{\left( {{x^ - }} \right)}^2} - {{\left( {{x_B}} \right)}^2}}
\right] + 2{{\left( {{x^
- }} \right)}^2}\left( {2{x^ - } + {x_B}} \right)}}f,\\
b &=&  - \frac{{\xi\eta\left( {{x^ + } - {x_B}} \right)\left[
{{{\left( {{x^ - }} \right)}^2} - {{\left( {{x_B}} \right)}^2}}
\right] + 2{{\left( {{x^ - }} \right)}^2}\left( {2{x^ + } - {x_B}}
\right)}} {{\xi\eta\left( {{x^ - } + {x_B}} \right)\left[
{{{\left( {{x^ - }} \right)}^2} - {{\left( {{x_B}} \right)}^2}}
\right] + 2{{\left( {{x^
- }} \right)}^2}\left( {2{x^ - } + {x_B}} \right)}}f,\\
c &=& \frac{{i\xi{x_B}\left( {{x^ + } + {x^ - }} \right)\left(
{{x^ + } - {x^ - }} \right)}} {{\xi\eta\left( {{x^ - } + {x_B}}
\right)\left[ {{{\left( {{x^ - }} \right)}^2} - {{\left( {{x_B}}
\right)}^2}} \right] + 2{{\left( {{x^ - }} \right)}^2}\left( {2{x^
- } + {x_B}}
\right)}}f,\\
d& =& \frac{{2i\eta{{\left( {{x_B}} \right)}^2}\left( {{x^ + } +
{x^ - }} \right)}} {{\xi\eta\left( {{x^ - } + {x_B}} \right)\left[
{{{\left( {{x^ - }} \right)}^2} - {{\left( {{x_B}} \right)}^2}}
\right] + 2{{\left( {{x^ - }} \right)}^2}\left( {2{x^ - } + {x_B}}
\right)}}f,\\
e& =& \frac{{\xi\eta\left( {{x^ - } + {x_B}} \right)\left[
{{{\left( {{x^ + }} \right)}^2} - {{\left( {{x_B}} \right)}^2}}
\right] + 2{{\left( {{x^ + }} \right)}^2}\left( {2{x^ - } + {x_B}}
\right)}} {{\xi\eta\left( {{x^ - } + {x_B}} \right)\left[
{{{\left( {{x^ - }} \right)}^2} - {{\left( {{x_B}} \right)}^2}}
\right] + 2{{\left( {{x^ - }} \right)}^2}\left( {2{x^ - } + {x_B}}
\right)}}f,
\end{eqnarray}
where $\xi$ and $\eta$ are arbitrary boundary parameters.

\section{Discussion}

By directly solving the reflection equation Eq.(\ref{rightbybe2}),
we obtain a boundary matrix with two free parameters  for the
$SU(1|1)$ spin chain model. Here, we only suppose the boundary
take the form like Eq.(\ref{rm}).  Whether there are other
boundary matrices, we need to explore further.

The  open  $SU(1|1)$ quantum spin chain with its right boundary
matrix satisfying reflection equation  Eq.(\ref{rightbybe1}) is
solved by analytical Bethe ansatz method\cite{26}.  It would be
interesting to investigate how to solve this  spin chain with its
boundary matrix satisfying Eq.(\ref{rightbybe2}).

\section*{Acknowledgement}
We thank W. Galleas for his helpful suggestions. This work is
supported by the keygrant project of Chinese Ministry of
Education(No.708082).


\begin{thebibliography}{99}


\bibitem{1y} Minahan J A and  Zarembo K, 2003 \textit{ JHEP} {\bf 03} 013 [hep-th/0212208]

\bibitem{1}   Beisert N and  Staudacher M, 2003 {\textit{Nucl. Phys.}} B  {\bf 670}
439 [hep-th/0307402]

\bibitem{2}  Belitsky  A V,  Derkachov S E,  Korchemsky G P and
Manashov A N, 2004 \textit{ Phys. Lett.} B {\bf 594}  385
[hep-th/0403085]




\bibitem{3}  Arutyunov G,  Frolov S,  Russo J and
 Tseytlin A A, 2003 \textit{ Nucl. Phys.} B {\bf 671}  3 [hep-th/0307191]


\bibitem{4}  Kazakov  V A,  Marshakov A,  Minahan J A and  Zarembo K,
2004 \textit{JHEP} {\bf 0405}  024 [hep-th/0402207]



\bibitem{5} Arutyunov G and  Frolov S, 2005 \textit{JHEP} {\bf 0502}
059 [hep-th/0411089]


\bibitem{6} Alday  L F, Arutyunov G and  Tseytlin A A, 2005 \textit{ JHEP} {\bf 0507}  002
[hep-th/0502240]

\bibitem{7}   Beisert N, 2007 \textit{ J. Stat. Mech.}  P01017 [nlin/0610017]

\bibitem{8}  Martins M J and  Melo C S, 2007  \textit{ Nucl. Phys.} B {\bf 785}  246 [hep-th/0703086]
\bibitem{8a}  Arutyunova G,  de Leeuw M,  Suzuki R and  Torrielli
A,  2009 [arXiv:0906.4783]

\bibitem{9}  Faddeev  L D,  1996 [hep-th/9605187]

\bibitem{10}   Sklyanin E K,  1988 \textit{J. Phys.} A {\bf 21} 2375

\bibitem{11}  Berenstein D and  Vazquez S E, 2005 \textit{JHEP} {\bf 06}  059 [hep-th/0501078]

\bibitem{12}  McLoughlin T and  Swanson I, 2005 \textit{Nucl. Phys.} B {\bf 723}  132 [hep-th/0504203]

\bibitem{13}  Agarwal A, 2006 \textit{ JHEP} {\bf 08} 027 [hep-th/0603067]



\bibitem{14}  Hofman D M, and  Maldacena J M, 2007 \textit{JHEP.} {\bf 0711}
00263 [arXiv:0708.2272]

\bibitem{15}  Ahn C and Nepomechie R I, 2008 \textit{JHEP.} {\bf 0805}
059 [arXiv:0804.4036]

\bibitem{16}   Beisert N, 2008 \textit{ Adv. Theor. Math. Phys.} {\bf 12} [hep-th/0511082]

\bibitem{17} Murgan R and  Nepomechie R I, 2008
\textit{JHEP} {\bf 0809} 085 [arXiv:0808.2629]

\bibitem{18}   Galleas W, 2009 \textit{Nucl.Phys.} B {\bf 820} 664
[arXiv:0902.1681]


\bibitem{19}   Nepomechie R I, 2009  \textit{JHEP} {\bf 0905} 100 [arXiv:0903.1646]

\bibitem{20}  Janik R A, 2006 \textit{ Phys. Rev.} D {\bf 73}
086006 [hep-th/0603038]

\bibitem{21}  Malara R,  Lima-Santos A 2006 \textit{ J. Stat. Mech.}
P09013 [arXiv:nlin/0412058 ]\\
 Lima-Santos A, Martins M J 2007 \textit{Nucl. Phys} B
{\bf 760} 184 [arXiv:nlin/0608063 ]


\bibitem{22} Lima-Santos A 2009 \textit{ J. Stat. Mech.} P04005 [arXiv:0810.1766 ]\\
Lima-Santos A 2009 \textit{ J. Stat. Mech.} P07045 [arXiv:0809.0421]\\
Lima-Santos A,  2009 \textit{ J. Stat. Mech.} P08006\\
Lima-Santos A, Galleas W 2008  [arXiv:0806.3659]



\bibitem{23} Guan X W 2000 \textit{J. Phys} A {\bf 33} 5391 [cond-mat/9908054]\\
Guan X W, Foerster A, Grimm U, Romer R A, Schreiber M 2001
\textit{Nucl. Phys} B {\bf 618} 650 [cond-mat/0106511]\\
Doikou A and Karaiskos N 2009 \textit{ J. Stat. Mech.} L09004
[arXiv:0907.3408 ]

\bibitem{24}  Murgan R,  Nepomechie R I, 2008 \textit{JHEP} {\bf 0806}
096 [arXiv:0805.3142]

\bibitem{25}  Beisert N, 2006 \textit{ Bulg. J. Phys.} {\bf 33S1}  371 [hep-th/0511013]

\bibitem{26} Nepomechie R I, Ragoucy E, 2008 \textit{JHEP} {\bf
0812} 025 [arXiv:0810.0515]





















\end{thebibliography}
\end{document}